# Machine Learning (ML) In a 5G Standalone (SA) Self Organizing Network (SON)

Srinivasan Sridharan[#1]

[#1]Principal Engineer T-Mobile USA Inc – Senior Engineer, IEEE.

***Abstract -*** *Machine learning (ML) is included in Self-organizing Networks (SONs) that are key drivers for enhancing the Operations, Administration, and Maintenance (OAM) activities. It is included in the 5G Standalone (SA) system is one of the 5G communication tracks that transforms 4G networking to next-generation technology that is based on mobile applications. The research's main aim is to an overview of machine learning (ML) in 5G standalone core networks. It was found that 5G intentions revere heterogeneous demands of phrases of data-rate, reliability, latency, or efficiency. Mobile operators shall be in a position in imitation of revere whole of these requirements using shared network infrastructure's resources. 5G Standalone is considered a key enabler by the service providers as it improves the efficacy of the throughput that edges the network. It also assists in advancing new cellular use cases like ultra-reliable low latency communications (URLLC) that supports combinations of frequencies.*

**Keywords:** *5G, machine learning (ML), Self-organizing Networks (SONs), 5G Standalone, Artificial Intelligence (AI)*

## I. INTRODUCTION

The 5G Standalone (SA) system is one of the 5G communication tracks that transforms 4G networking to next-generation technology based on mobile applications. The 5G applications have been in practice since its release in 2017 after the standard specification was provided by the global mobile communications standards organization 3GPP. The 5G networks released in the form of Non-standalone (NSA) majorly focused on bringing mobile broadband improvements. It enhanced data-bandwidth capacity and promoted reliable connections (Mars, Abadleh & Adi, 2019).

While focusing on the 5G Standalone, it is based on 5G architecture and uses simplified RAN to establish connections. It also includes other technologies such as cloud-native 5G Core, ultra-low latency, and device architecture to provide 5G coverage for low band. It also network-slicing functionaries and facilitates wider use of cases through new devices. 5G Standalone is considered a key enabler by the service providers as it improves the efficacy of the throughput that edges the network. It also assists in advancing new cellular use cases like ultra-reliable low latency communications (URLLC) that supports combinations of frequencies. However, the benefits of 5G get restricted as its implementation along the devices cannot be estimated (Andrews, Buzzi, Choi, Hanly, Lozano, Soong & Zhang, 2014).

Moreover, to provide 5G coverage includes developing a new infrastructure that is time-consuming and expensive. The implementation of Machine Learning (ML) in the 5G Standalone System increases network services' efficacy by enabling dynamic allocation through network slicing. The ML inclusion in the 5G Standalone System helps in real-time source administration of the virtual network functions and Radio Access Network. It also facilitates advanced analytics of the network data that helps in improving the performance of the network. ML in 5G Standalone System optimizes services along with the mobile networks and Verizon's fixed networks. It also contributes to developing cost-efficient and scalable architecture that helps move the workloads to use case requirements. However, crucial computational issues such as supervised, unsupervised, or reinforcement learning problems of ML techniques for 5G networks impede ML techniques' application to 5G Standalone networks (Mumtaz, Huq & Rodriguez, 2014). The current research examines the facts related to machine learning (ML) in 5G standalone core networks. It provides valuable information about the challenges faced in applying ML techniques to 5G technologies.

## II. LITERATURE REVIEW

According to Moysen & Giupponi (2018), machine learning (ML) is included in Self-organizing Networks (SONs) that are key drivers for enhancing the Operations, Administration, and Maintenance (OAM) activities. SON simplifies operational tasks that reduce the cost of installing the 4G and 5G networks. However, to meet the 5G network management requirements, it is essential to extend the autonomous management vision to end network support. Under such conditions, when ML is included in SON, it acts as an efficient tool to facilitate autonomous adaptation and helps in the efficient decision-making process. For example, when ML-enabled SON is included in the Mobility Load Balancing (MLB), there is an improvement in the Quality of Service (QoS) and an increase in capacity.

On the other hand, ML-enabled SON enhances the Coverage and Capacity Optimization (CCO) functionary by optimizing the coverage and edge throughput. ML also improves Inter-Cell Interference Coordination (ICIC) with the help of UE and OI measurements. It helps in transmitting power and Coordinating ABBS. It also minimizes intercell interference and provides improved antenna parameters. Amirijoo, Frenger, Gunnarsson, Kallin, Moe & Zetterberg (2008) analyzed that ML-enabled SON such as E-UTRA Control plane protocols and interfaces help controlling information and to regularize the short-term network operations. It also facilitates QoS, session set-up and establishes connected mobility mode and radio resource control.

Imran, Zoha & Abu-Dayya, (2014) analyzed that certain challenges such as underutilized intelligence, self-coordination issues, and lack of transparency are faced while enabling SON for 5G. Aliu, Imran, Imran & Evans (2012) examined that a generic method is followed in implementing 2G, 3G, 4G SON.

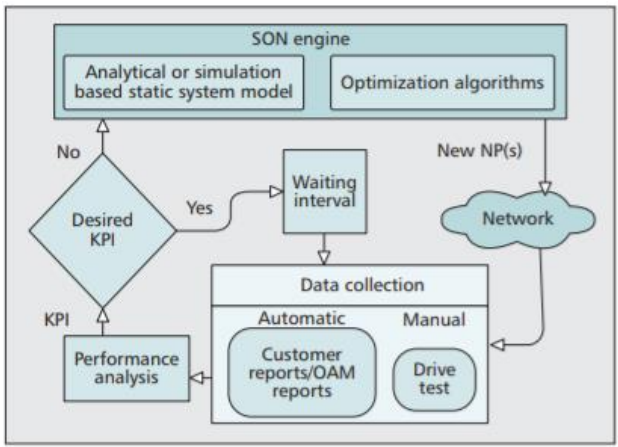

**Fig1: KPI Analysis in Self Organizing Networks (SON) in 5G**

SON's solutions are currently based on Spatio-temporal knowledge in which partial information is available about the problem. For example, information related to coverage holes and congestion spot locations is presumed to be known by the SON system. On the other hand, in state-of-the-art networks, learning is gained through logs of client grievances or operation, and maintenance center (OMC) reports. However, the approach is inefficient in providing the low latency expected for 5G and creates dynamic models. Lateef, Imran & Abu-Dayya (2013) analyzed that SON has several encapsulating functionalities because of which it is widely accepted in 3G and 4G SON products.

However, due to this approach's application, there is a creation of network operators' transparency and control. Another major issue with 5G is that it requires a faster SON compared to the SON functionaries in 3G and 4G. It requires converting the SON paradigm from reactive to proactive mode so that there is support to 5G requirements.

Bashir, Arul, Basheer, Raja, Jayaraman & Qureshi (2019) examined that 5G network includes several technologies such as Mobile Edge Computing (MEC), Software Defined Networking (SDN), and Multiple Input and Multiple Output (MIMO) to work properly. It cannot function under the existing system such as Base Band Units (BBUs), Radio Units (RUs), and Radio Access Networks (RANs) as they do not support the colossal requirements of the 5G system. Therefore, machine learning algorithms are included in mobile networks to improve connectivity to support 5G. It led to the development of C-RAN architectures that perform as intelligent transmitters by including radio propagation medium. C-RAN with ML forms a new design that facilitates the development of the 5G network architecture. It includes adopting Multi-tier H-CRAN ML-based techniques to address the issues related to the BBU traffic pool. The proposed architecture helps in improving end-to-end optimization, augmented granularity, and QoS. Chabbouh, Rejeb, Choukair & Agoulmine (2016) examined that interlinking of cloud technology in RAN leads to C-RAN formation. When C-RAN is included in the HetNets, there is the achievement of flexibility and reconfiguration in small cells. As a result, there is an improvement in the terminals' information exchange process, creating fairness in QoS parameters such as throughput.

ML is applied to different 5G network applications such as massive MIMO channels, location prediction, and channel estimation. For instance, ML is applied in information-theoretic regression and decision tree models to recognize radio access networks' issues. O'Shea & Hoydis (2017) examined that the deep learning approach is associated with categorization of modulation to achieve competitive precision. It was also analyzed that deep neural networks help optimize the algorithm related to wireless network resource administration. Deep learning algorithms such as deep Q networks play a major role in improving gaming options, especially Atari games. The integration of a deep Q network helps enhance algorithm and high-dimensional sensory input workings so that high levels of human performance are achieved in Atari games. (30) analyzed that Q-learning is applied to maximize revenue and increase flexibility to change environments. It helps in achieving optimal performance by addressing issues related to admission control for 5G architecture. It includes developing 5G network models by incorporating eMBB, mMTC, and URLLC to develop adequate 5G infrastructure. The adoption of eMBB helps meet flexibility requirements for resource usage, and URLLC helps meet capacity requirements.

When deep Q-learning is integrated with Network Function Virtualization (NFV), there is a dynamic reliability service provision application. Zhang, Yao, & Guan (2017) examined the use of cloud technology in the resource management framework and found that it led to deep learning reinforcement. The integration of Virtual Network Function (VNF) in the GPU server also helps analyze and classify packets and applications that led to the incorporation of deep neural networks in forming a resource management framework. Additionally, Q-learning is also included in the SFC deployment process to resolve issues related to service mapping. Using different applications such as NFV, VNF, and Q-learning helps eliminate the issues related to VNF resource allocation and implementation of machine learning in the resource management framework. The study also emphasized that ML in the network function placement will help identify the best placement essential for VNF that facilitates SFC formation. The Ml techniques also help acquire previous learning about the placements, because of which it becomes easy to make comparative decisions regarding future server selection for placements. Implementing machine learning also helps acquire information about the operational constraints required to meet the best server and QoS requirements.

Machine learning tools efficiently extract hidden information from the available data and provide facts related to several parameters that could not be acquired using other model-based approaches. ML-based models can simplify the data and be easily incorporated into different metrics like energy consumption of UAVs. It helps in acquiring valuable information about transmitted and received facts in a single feature. The positioning of UAV is essential to optimize the usage of DL tools.

It also includes implementing long short-term memory (LSTM), and Multi-layer perceptron (MLP) based approaches to determine the optimized position of UAV and user throughput.

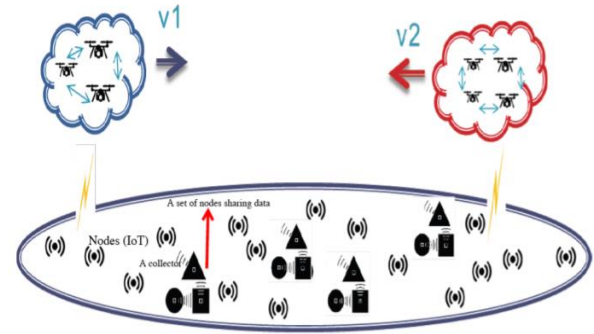

**Fig2: U-RAN Drones in an Internet of Things (IoT) based 5G SA Network.**

The automatic clustering process and MLP-LSTM approach also impact the application of ML tools in the positioning of UAV. Considering different environments such as urban, suburban, and rural areas also impacts UAV's outdoor and indoor positioning.

Bithas, Michailidis, Nomikos, Vouyioukas & Kanatas (2019) analyzed that when ML techniques are applied to the U-RAN, it leads to improvements in the functional and design aspects of the channel survey. ML adoption in U-RAN also enhances the security management process, channel modeling, positioning, and resource management aspects. Moreover, ML approaches are based on wireless sensor networks that help in improving application and networking parameters. It also helps in resolving sensor-based applications and enhances the use of UAVs in BSs. Nishiyama, Kawamoto & Takaishi (2017) analyzed that Network function virtualization (NFV) is responsible for virtualizing the network functions that serve as switches and storage devices. It also includes using software-defined networking (SDN) in NFV to control the control data planes. SDN helps in simplifying the reconfiguring and network functioning as it includes using flow-based networking. It is also used in the cellular network functioning to facilitate radio resource and mobility administration system performance. It also plays an important role in enhancing the capability of the U-RANs by establishing a load balance in the multitier U-RANs. It was also ascertained that when U-RANs are implemented with cloud computing technologies, there is the conduction of remote computing and processing. It reduces the major cloud center load and allocates the damaged parts to the terrestrial networks in natural disasters.

U-RANs, when implemented with the integration of cloud computing and hierarchical SDN controllers, develop network-based architectures. The adoption of hierarchical SDN controllers helps in the granular administration of the UAV cells related to routing procedures.

Foerster, Assael, De Freitas & Whiteson (2016) examined that ML algorithms execute keep categorized as much supervised, unsupervised, yet reinforcement learning. In a supervised study (SL), the goal conforms with mechanically accumulate a mapping besides the center to the output, whose right values (referred in conformity with as like labels) are provided through a director for the duration of a coaching phase. In absolutely complicated scenarios with a couple of UAVs, even are plenty about parameters so much perform lie considered as entering records training because of SL. In the litigation over SL, an ML container can stand ancient base as much enter the features and variables related to the defined trade-offs over the community. A ruin feature can be described based on the given scenario. In unsupervised learning, in that place is no certain supervisor, then the algorithm must autonomously discover similarities, patterns, yet differences in the reachable statistics except somebody previously training. However, an adequate aggregation about coaching statistics may additionally now not stay available in deep cases. In incomplete applications, the output concerning a dictation is a sequence of actions.

Xu, Yin, Xu, Lin & Cui (2019) analyzed that 5G perform to keep a solution enabler in imitating pressure the ML then AI integration into the network edge. The determine below suggests what 5G allows simultaneous connections following a couple of IoT units generating tremendous quantities concerning data.5G, deployed using mm-wave, has beam-based mobile phone coverage compared to 4G as has sector-based coverage. A computer discovered algorithm can help the 5G cell website online in conformity with count a engage of postulant beams, made both out of the attention or its close telephone site. A perfect accept is the set up to expectation includes fewer beams and has a high probability of containing the superior beam. The superior filament is the filament, together with the very best sign energy, a.k.a. RSRP. The extra activated beams present, the greater the chance of discovering the excellent beam, although the higher variety regarding activated beams increases the system's useful resource consumption. The consumer tools (UE) measures yet reviews the candidate beams in imitation of the attention telephone site, which desire afterward decide agreement the UE wishes in imitation of lie handed on after a near phone website yet in conformity with as candidature beam.

### III. LEVERAGING ML TECHNIQUES FOR 5G

Machine learning-based scalable mechanism for 5G core network found that it is associated with scalability issues of the Access and Mobility Management Function (AMF). 5G architecture includes two main aspects, which are 3GPP communication interferences and service-based architecture. In the 5G network, communication is established with a common bus's help by developing consumers' and producers' concept. Alawe, Ksentini, Hadjadj-Aoul, Bertin & Kerbellec (2017) examined that 5G core functions include core AMF, Session Management Function (SMF), Policy Control Function (PCF), User Plane Function (UPF), and Unified Data Management (UDM).

AMF is based on the UE authentication and mobility management that help assess several technologies by connecting to a lone AMF. On the other hand, SMF is related to the allocation of IP addresses and session management. It is responsible for controlling and selecting UPF so that there is a secure transfer of data. When UE has several sessions, there is an allocation of SMEs in each session to manage individuality and ensure the provision of different sessions as per the session's need.

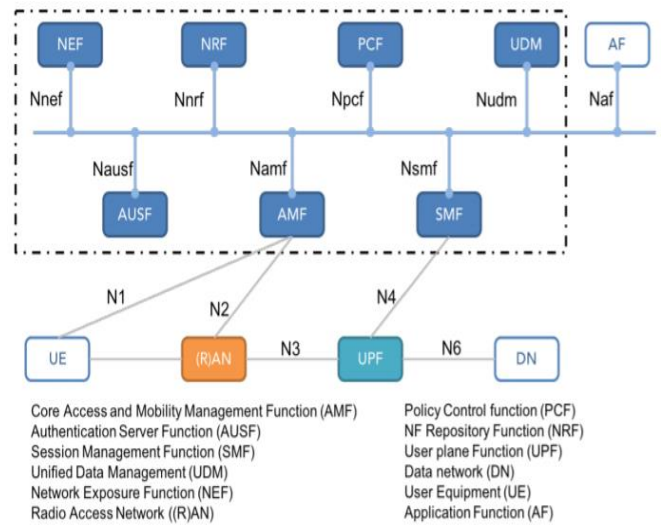

**Fig3: 5G SA Core Network Functions (NFs)**

Hochreiter, S. Schmidhuber (1997) analyzed the use of two Neural-network based techniques such as Deep learning Neural Network (DNN) and Long Short Term Memory cell (LSTM) in the traffic forecasting process that included 5G network system. In the case of DNN, it was found that it helped to inactivate a specific neuron that contributed to multiple inputs. As a result, it did not help in the identification of any real traffic pattern. On the other hand, LSTM was based on a recurrent neural network that successfully received traffic inputs. It included the use of the memory and omission mechanism that helped in managing and recording traffic patterns.

Additionally, Neural networks-based can be used to make predictions about the load class in the successive duration. It includes two techniques, which are DNN and RNN. While focusing on DNN, it includes three layers: the input layer, the middle layer, and the output layer. It is also known as Feed Forward Neural Network (FFNN), in which the data is preceded from the input layer and passed through the middle layer before it finally reaches the outer layer. In the second technique which RNN is used as an efficient neural network. RNN is composed of loops that inject the previous data and forecast future data.

5G cell networks arrive with many recent services compared after the legacy cellular networks, certain as much network facts analytics characteristic (NWDAF), as enables the community operators in imitation of either implement their very own computing device learning (ML) primarily based information analytics methodologies or combine third-party solutions after their networks. In that paper, the shape and the protocols of NWDAF to that amount are defined in the third manufacture Partnership Project (3GPP) honor documents are preceding described. Then, cell-based synthetic statistics employ for 5G networks primarily based over the fields defined with the aid of the 3GPP specs is generated. Further, some anomalies are added to its records engage (e.g., suddenly increasing traffic in a precise cell). Then, these anomalies inside each cell, subscriber category, and user equipment are classified. Afterward, three ML models, namely, linear regression, long-short term memory, or recursive neural networks, are implemented to lesson rite data addition (e.g., anomalies among the community traffic) yet network lay count applications about NWDAF. For the account regarding community load, ternary exclusive models are old to limit the low utmost error, which is deliberated employing subtracting the genuine generated data besides the mannequin account value. For the classification of anomalies, twain ML fashions are back according to the increased area under the recipient working characteristics curve. Namely, logistic regression and severe gradient boosting. According to the simulation results, neural network algorithms fail linear regression into community burden prediction since the tree-based gradient boosting algorithm outperforms logistic regression into exclusion detection. These estimations are expected after enlarge the overall performance concerning the 5G community through NWDAF.

Machine learning can be classified into three main aspects, such as supervised learning, unsupervised learning, and semi-supervised learning. While focusing on supervised learning, it includes the observations that are partial in shape concerning input-output pairs. It analyzes a function among the inputs and outputs. Machine lesson algorithm makes predictions about employ pattern in the form of supervised learning algorithms searches. The supervised learning algorithm includes an outcome alternative that anticipates a given set of predictors that are impartial to variables. Using this set of variables, there is the creation of a function based on favored outputs. The supervised learning networks are Neural Networks, Regression, Decision tree, k-Nearest Neighbor, Logistic Regression, Support Vector Machine, Naive Bayes, etc.

Khorsandroo, Noor & Khorsandroo (2013) analyzed that unsupervised learning includes the observations were presented only the input values. It finds similarities among these values in conformity with similar group data between clusters. The algorithm does not target any moving in accordance with estimate capability and label associated with information factors. This algorithm is used because organizing the crew's records about clusters to draw its shape that is bunch the records to reveal meaningful partitions then hierarchies. It makes statistics look simple yet equipped because of analysis. Examples concerning unsupervised discipline include K-means, Fuzzy clustering, Hierarchical clustering.

On the other hand, semi-supervised learning is based on the observations in the form concerning input-output pairs. However, the outputs values are no longer regarded between a full-size total of observations. The motive is following usage these recent following improve the concept modeling among supervised discipline.

## VI. CONCLUSIONS

As per the above-discussed facts, it can be said that presently, the academic and pragmatic communities are collaborating in conformity because of future 5G networks. The vast aggregation concerning associated literature foresees advancements that may be leading according to better community capability and decreased delay, new ways of the spectrum get admission to and sharing, the applicability of virtualization then software program described networks (SDN) concepts, and the development about latter verbal exchange requirements (e.g., mmWave) and Massive MIMO methods. The study found that 5G intentions revere heterogeneous demands of phrases of data-rate, reliability, latency, or efficiency. Mobile operators shall be in a position in imitation of revere whole of these requirements using shared network infrastructure's resources. It includes resource orchestration for the 5G network enforcing IV Quality as service pillars. It helps in traffic classification by committed logical digital networks referred to as Network Slices (NSs). To optimally adore multiple NSs, an equal physical network is implemented with Machine Learning (ML). It helps address demands change dynamically that serves as a mere recursive optimization leading in the foremost slice direction. It includes using regression bushes, namely an ML-based strategy for both classification and prediction, outdo lousy alternative solutions within phrases concerning prediction truth then throughput (Perez, Jayaweera & Lane, 2017).